\documentclass[9pt,sigconf,letterpaper]{acmart}
\usepackage[english]{babel}
\usepackage{blindtext,xcolor}

\usepackage[utf8]{inputenc}
\usepackage{microtype}
\usepackage{wasysym}

\usepackage{enumitem}
\usepackage{graphicx}
\usepackage{array}
\usepackage{balance}
\usepackage{siunitx}
\usepackage{booktabs}
\usepackage{xparse}
\usepackage{tabularx}
\usepackage{url}
\usepackage[font=small]{caption}
\usepackage{subcaption}
\usepackage{makecell}
\usepackage{xspace}
\usepackage{multirow}
\usepackage{fancyvrb}
\usepackage{wrapfig}

\newcommand{\ZDNS}{ZDNS\xspace}
\newcommand{\zdns}{\ZDNS}
\newcommand{\zdnss}{\ZDNS's\xspace}
\newcommand{\dns}[1]{{\small \texttt{#1}}}

\acmYear{2022}\copyrightyear{2022}
\acmConference[IMC '22]{ACM Internet Measurement Conference}{October 25--27, 2022}{Nice, France}
\acmBooktitle{ACM Internet Measurement Conference (IMC '22), October 25--27, 2022, Nice, France}
\acmPrice{15.00}
\acmDOI{10.1145/3517745.3561434}
\acmISBN{978-1-4503-9259-4/22/10}

\begin{document}

\title{\zdns: A Fast DNS Toolkit for Internet Measurement}

\settopmatter{authorsperrow=4}

\author[]{Liz Izhikevich}
\affiliation[]{Stanford University}

\author[]{Gautam Akiwate}
\affiliation[]{Stanford University}

\author[]{Briana Berger}
\affiliation[]{Stanford University}

\author[]{Spencer Drakontaidis}
\affiliation[]{Stanford University}

\author[]{Anna Ascheman}
\affiliation[]{Stanford University}

\author[]{Paul Pearce}
\affiliation[]{Georgia Institute of Technology}

\author[]{David Adrian}
\affiliation[]{Stanford University}

\author[]{Zakir Durumeric}
\affiliation[]{Stanford University}

\renewcommand{\shortauthors}{L. Izhikevich et~al.}


\begin{abstract}

Active DNS measurement is fundamental to understanding and improving the DNS ecosystem. However, the absence of an extensible, high-performance, and easy-to-use DNS toolkit has limited both the reproducibility and coverage of DNS research. In this paper,
we introduce \zdns, a modular and open-source active DNS measurement framework optimized for large-scale research studies of DNS on the public Internet.
	We describe \zdnss architecture, evaluate its
	performance, and present two case studies that highlight how the tool can be used to shed light on the operational complexities of DNS\@.
	We hope that \zdns will enable researchers to better---and in a more reproducible manner---understand Internet behavior.

\end{abstract}

\maketitle

\section{Introduction}

The Domain Name System (DNS) plays a critical role on the Internet, from acting as the phone book of the web to controlling traffic routes for major content delivery networks~\cite{akamai,cloudfront} and authenticating services~\cite{rfc6698}. DNS has proven to be complex to manage, which, paired with its ubiquity, has led to major Internet outages~\cite{akamai_outage,aws_outage,fastly_outage} and the hijacking of prominent services~\cite{talos_sea_turtle_2019,krebs_talos_2019}. 
With billions of names, millions of resolvers, and dozens of types of records defined across hundreds of RFCs, DNS has become a massively complex and distributed ecosystem that is not fully visible to researchers.
Further inhibiting visibility, DNS behavior is often hidden behind recursive and caching resolvers.

While there exist many tools for actively querying DNS, none expose internal DNS operations (e.g., responses from each step of the recursive process) while scaling to today's namespace and providing the extensibility needed for quickly answering new types of research questions. Consequently, researchers frequently resort to building their own specialized scanning solutions~\cite{akiwate2020unresolved,mao2022assessing,ccr2018_caa_first_look}, which are expensive and error prone to develop. Further, most tools have remained closed source, hampering reproducibility.

In this work, we introduce \zdns, an open source measurement toolkit for large-scale active DNS research. 
\zdns is composed of: (1)~a DNS library that exposes the internal characteristics of DNS operations, (2)~a core framework that isolates and orchestrates non-DNS specific scanning logic, and (3)~composable modules that allow researchers to easily add new DNS queries and records, including 65~already-implemented record types.
Critically, \zdns implements its own internal caching and recursion, which is key to exposing internal DNS operations and querying a large number of {unique} names. 
\zdns provides a simple command line interface for scanning and outputs results in programmatically interpretable JSON\@. 

We evaluate \zdns and show that it successfully resolves 85~times more domains per second than prior work~\cite{dig}, performing 90K~lookups per second when using an external recursive resolver. 
\zdns resolves 50M~domains in 10~minutes and queries the PTR records of the full public IPv4 address space in 12~hours.
When performing its own internal recursion, \zdns exposes all recursive details to the user and resolves 50M~domains in 46~minutes and 100\% of the public IPv4 address space in 116~hours.
We supplement our evaluation with two case studies that explore (1) the customizability and extensibility of \zdns when exposing internal DNS operations to analyze redundant nameserver deployment, and  (2) the versatility of \zdns and the importance of accounting for blind spots in the coverage of open data sets.

Since its open source release, \zdns has been used by a series of Internet measurement papers~\cite{pearce2017global,fainchtein2021holes,patil2019can,wang2019blacklist,maroofi2021adoption,patil2020privacy,fanou2020unintended,maroofi2020defensive,dimova2021cname,ruth2018first,barman2022not,ramesh2020decentralized,mirian2018https,martiny2022privacy,dorey2017internet} and serves as the foundation for several open data sets~\cite{caida_dset,kountouras2016enabling,censys-2015}. 
Given its widespread use and now stabilized codebase, we formally introduce \zdns to the community in order to promote an awareness and full understanding of the tool's motivation, architecture, performance, capabilities, and caveats.
High-performance, open source, Internet-wide scanners (e.g., ZMap~\cite{zmap-2013}, ZGrab~\cite{censys-2015},  Masscan~\cite{graham2014masscan}) have already helped break down the barrier of scalability and reproducibility for hundreds of research papers that study the Internet ecosystem; we hope that \zdns will do the same for DNS\@.
\zdns is released under the Apache 2.0 licence at \url{https://github.com/zmap/zdns}.



\section{Related Work}

Active DNS measurement has been critical to understanding the operation~\cite{patil2019can,fanou2020unintended,ruth2018first,ramesh2020decentralized}, privacy~\cite{patil2020privacy,fainchtein2021holes,dimova2021cname}, and security~\cite{pearce2017global,wan2020origin,maroofi2021adoption,maroofi2020defensive} of the DNS ecosystem. There exist a wide range of tools from dig~\cite{dig} to MassDNS~\cite{massdns}---a high-performance stub resolver that was developed in parallel to ZDNS\@. Often, these tools are paired with a public resolver like Cloudflare's \texttt{1.1.1.1}~\cite{cloudflare_res} or Google's \texttt{8.8.8.8}~\cite{google_res}, or an external recursive resolver like PowerDNS~\cite{powerdns}, Unbound~\cite{unbound}, or Bind~\cite{bind}. We evaluate these tools in Section~\ref{sec:eval}. 

However, existing tools do not always extend to the research question at hand. For example, MassDNS is not engineered to perform iterative lookups and expose the DNS lookup chain. Unbound implements only features based on RFC standards~\cite{unbound_rfc}, resulting in over 30~feature requests that have been unresolved for years~\cite{unbound_issues}.
Consequently, researchers regularly develop their own purpose-built measurement tools~\cite{mao2022assessing,akiwate2020unresolved,ccr2018_caa_first_look}.
These independent scanning solutions are not always optimized for performance---frequently relying on sub-sampling the DNS/rDNS search-space~\cite{richter2016beyond,fiebig2017something}---and are not always public~\cite{mao2022assessing,akiwate2020unresolved,ccr2018_caa_first_look}, therefore hindering reproducibility and instigating an unnecessary reinvention of the same tool.

The measurement community has also produced datasets built on top of active DNS querying, including  OpenINTEL~\cite{openintel}, ActiveDNS~\cite{kountouras2016enabling}, Rapid~7~\cite{rapid7}, Farsight~\cite{farsight}, and Censys~\cite{censys-2015}. More than 100~papers have relied on these datasets to understand~\cite{sommese2020parents,ccr2018_caa_first_look} and protect~\cite{jonker2017millions,chung2017understanding} DNS infrastructure.
While these open data sets have considerably benefited the community,
we demonstrate in Section~\ref{sec:caa}, along with prior work~\cite{fiebig2018rdns}, that the selective coverage of data sets often requires supplemental active measurement (e.g., measuring additional zones or response inconsistencies). 
Notably, ActiveDNS and Censys are built on top of \zdns.



\section{\zdns Architecture}
\label{sec:arch}

In this section, we describe \zdnss architecture and guiding principles that inform its extensible and performant design.
To support reproducibility, \zdns is open source and can be found at \url{https://github.com/zmap/zdns}.

\subsection{Requirements and Guiding Principles}

We built \zdns because there were no readily available tools for quickly performing millions of unique DNS queries. The following observations, goals, requirements, and lessons learned from other open source measurement tools, inform our architecture:

\vspace{3pt} \noindent \textbf{Internal Recursion.}\quad Recursive resolvers
hide many characteristics of DNS operations (e.g., lame and dangling delegations~\cite{akiwate2020unresolved}). Further, public
recursive resolvers often rate limit queries~\cite{googleRateLimit}---cloud providers like 
Amazon have explicitly asked that we not use their infrastructure for large scale experiments---and open source resolvers like Unbound~\cite{unbound} are not easily extensible. 
\zdns must be able to perform its own recursive resolution and expose lookup chains in addition to supporting external recursive resolvers. 

\vspace{3pt} \noindent \textbf{High Performance.}\quad  DNS experiments frequently require querying a large number of names. There are 1.5B~unique FQDNs in public
Certificate Transparency logs~\cite{censysSearch}, 161M domains in the Verisign \texttt{.com} zone
file~\cite{verisign}, 6M recursive resolvers on the public Internet~\cite{censysSearch}, and collecting
reverse PTR records for all of IPv4 requires querying 3.7B~publicly accessible IPv4
addresses. \zdns needs to be performant enough to lookup PTR records
for all IPs in a few days and known names in a few hours to keep up with regular Internet churn.

\vspace{3pt} \noindent \textbf{Safe.}\quad The DNS protocol is
defined across at least 285~RFCs and there are more than 65~types of DNS records in 2022.
Internet servers regularly respond with malformed responses due to misconfiguration and intentionally malicious
operators. \zdns needs to be written in a memory-safe language and support a
modular interface where researchers can easily {and} safely implement new
functionality. 

\vspace{3pt} \noindent \textbf{Extensible.}\quad Complex, difficult-to-use tools lead to measurement error because there is no ground truth against which to compare results. Similarly, operators will not use tools with a high barrier to entry, instead opting to build other tools, which may not provide the same level of accuracy. \zdns must be easy to understand and use. This precludes any complex, distributed setups and dictates that \zdns be easy to install and collect correct results.

\subsection{Architecture}


\zdns architecture is composed of three primary components: (1) a DNS library that fully exposes DNS lookup chains, (2) a framework for easy command-line interaction, and  (3) composable modules that facilitate easy extensibility. We detail each below:

\vspace{3pt}
\noindent \textbf{DNS Library.}\quad 
\zdns implements its own caching recursive resolver library on top of 
Miek Gieben's DNS implementation~\cite{miekg}. The \zdns library provides recursive lookups, caching, validation,
exposed transcripts of exchanged packets and easy DNS question construction
and answer parsing.
It also
supports using an external recursive resolver and various performance optimization (e.g., UDP socket reuse), which we discuss in Section~\ref{sub:sec:perf_opt}.
\zdns modules are given direct access to the DNS library, eliminating the need to redundantly implement DNS query logic for varying DNS queries/records.  

\vspace{3pt} \noindent \textbf{Framework.}\quad
The core framework is responsible for facilitating command-line configuration, spawning
lookup routines, delegating control to modules, encoding, decoding and
routing data to/from routines, load balancing against upstream resolvers,
aggregating run-time statistics, and ensuring consistent output. 
The framework is light-weight, accounts for only a quarter of the \zdns code
base, and is absent of most DNS-specific logic. 

\vspace{3pt} \noindent \textbf{Modules.}\quad
Modules are responsible for providing DNS query-specific logic, including logic specific to a DNS query or record, expected query response, additional command line flags, global and per-routine initialization.
Developers add modules by implementing a Go interface
that defines a \texttt{DoLookup} function that accepts two parameters: (1) a
name to be queried and (2) a name server. \texttt{DoLookup} returns an \texttt{interface{}} with
JSON export tags (i.e., a struct with tags that hint to our framework how to
label the fields when converting the struct to JSON). 
As we later discuss, because the \zdns framework orchestrates concurrency through
light-weight routines, \texttt{DoLookup} functions can be simple (e.g.,
open a UDP socket, create and send packet, wait for response or socket timeout,
close socket, and return any relevant data from the response packet). We show an example 
\texttt{DoLookup} function and a complete module in Appendix~\ref{app:sub:ex_mod}.

\vspace{3pt}
\noindent
By having an existing core framework and helper library, many
simple modules (e.g., \texttt{CAA} query) are implemented in a few lines
of code that simply set the query type and return the resulting DNS answers.
\zdns modules support both recursive and iterative
lookups, and can be used to query a single resolver for a large number of names,
a large number of servers for a single name, or a combination of both. 

\subsection{Implemented Modules}

\zdns currently includes three types of modules:

\vspace{3pt} \noindent \textbf{Raw DNS modules.}\quad Basic modules provide the raw DNS response from
a server similar to dig, but as structured JSON records. There exist modules for most types of DNS records today.\footnote{We support querying for and parsing these types
of records: A, AAAA, AFSDB, ANY, ATMA, AVC, AXFR, CAA, CDNSKEY, CDS, CERT,
CNAME, CSYNC, DHCID, DMARC, DNSKEY, DS, EID, EUI48, EUI64, GID, GPOS, HINFO,
HIP, ISDN, KEY, KX, L32, L64, LOC, LP, MB, MD, MF, MG, MR, MX, NAPTR, NID,
NINFO, NS, NSAPPTR, NSEC, NSEC3, NSEC3PARAM, NXT, OPENPGPKEY, PTR, PX, RP,
RRSIG, RT, SMIMEA, SOA, SPF, SRV, SSHFP, TALINK, TKEY, TLSA, TXT, UID, UINFO,
UNSPEC, and URI.}

\vspace{3pt} \noindent
\textbf{Lookup modules.}\quad Raw DNS responses frequently do not provide the data users need. For example, an \texttt{MX} response may not include the associated \texttt{A} records in the additionals section, requiring an additional lookup. To address this gap and provide a friendlier interface, we also provide several alternative lookup modules: alookup and mxlookup. mxlookup will additionally do an \texttt{A} lookup for the IP addresses
that correspond with an exchange record. alookup acts similar to nslookup and
will follow CNAME records.

\vspace{3pt} \noindent
\textbf{Misc modules.}\quad Misc modules provide other alternative means of querying
servers, such as extracting the version of resolvers  (e.g.,~\texttt{bind.version}).

\subsection{Performance Optimizations}
\label{sub:sec:perf_opt}

While \zdnss architecture facilitates extensibility, several
optimizations are critical to its performance.

\vspace{3pt} \noindent \textbf{Parallelism.}\quad
The majority of clock time expended making a DNS query is spent waiting for a
response rather than constructing or parsing DNS packets. Thus,
efficiently parallelizing a sufficiently large number of concurrent queries is crucial to achieve the performance we need.
Inspired by ZGrab~\cite{censys-2015}'s success---a popular open source high-performance application layer scanner written in Go---, we build \zdns in Go such that we can utilize the language's ability to efficiently manage thousands
of concurrent queries using lightweight routines. Go is memory-safe and
garbage collected, which facilitates providing a safe but extensible platform
while remaining highly performant. 
While \zdnss architecture is similar to ZGrab, the ratio of waiting to work is much higher than for application
handshakes which often require expensive cryptographic operations or parsing
large amounts of data. As we discuss in the next section, optimal performance
requires around 50K~concurrent queries---about 5--10~times more than ZGrab---which introduces new challenges. 

\vspace{3pt} \noindent \textbf{UDP Socket Reuse.}\quad Creating a socket for every lookup is exorbitantly expensive because each socket is used to
send and receive only two packets before being torn down and recreated. 
Instead, we establish and maintain a single long-lived raw UDP socket in each
lightweight routine for the lifetime of the program execution. Raw UDP sockets
bind to a static source port, and can be used to send UDP packets to arbitrary
destination IP/ports. This eliminates per-connection socket overhead,
without requiring us to manually construct IP and Ethernet headers for each
request. To support more threads than the size of the OS ephemeral port range,
we support binding to multiple IP addresses. This approach does not accommodate
TCP queries (e.g., truncated records) but we find that this rarely
occurs in practice (e.g., when querying the A records of a random sub-sample of 10~million domain names found in the Censys certificate transparency logs, 0.4\% responses return truncated).
Thus, \zdns by default only establishes TCP connections as needed.
Nevertheless, \zdns can optionally be configured to scan using only TCP\@. 

\vspace{3pt} \noindent \textbf{Selective Caching.}\quad While popular recursive
resolvers like Unbound cache results in order to quickly respond to lookups for
frequently queried names, we expect \zdns users to frequently lookup unique
names.  Caching remains critical to avoid repeating initial queries in iterative
resolutions (e.g., repeatedly querying root resolvers for \texttt{.com}'s name
servers), but caching the results for the names queried causes unnecessary
thrashing. Thus, we selectively cache only \texttt{NS} and glue records to
help with future recursion, but do not cache any results for the leaf names being
directly queried.

\vspace{3pt} \noindent \textbf{Increased Garbage Collection.}\quad Decreasing the
frequency of garbage collection is typically associated with improved
performance. However, the opposite is true for \zdns. Quadrupling the frequency that garbage collection occurs increases the
throughput, likely because short collections can be
interspersed  between other request processing and do not cause 
connections to timeout while waiting for garbage collection to finish.

\vspace{3pt} \noindent As we detail in the next section, our optimizations allow
\zdns to consistently perform over 90K~lookups per second across billions of consecutive lookups. 

\subsection{Ethics}
Any open source scanner can be used by researchers to understand and protect the Internet, and abused by adversaries to find vulnerabilities. 
We design \zdns with the ability to perform its own recursion internally in order to avoid overloading public recursive resolvers.
\zdns parameters allow the user to directly modulate the number of lookups-per-second, thereby minimizing packet drop and the unnecessary overloading of external networks and servers. 
We include an extensive README with \zdns that recommends users to coordinate with their upstream provider before performing high-volume scans. 
When executing scans and developing \zdns, we follow the community standards for good Internet citizenship outlined by Durumeric et al.~\cite{zmap-2013}.


\section{Evaluation}
\label{sec:eval}

We evaluate \zdnss scalability, execution time, and success rate when performing billions of queries and compare \zdns to a set of existing tools. We show that \zdns queries the PTR records of all IPv4 addresses in 12~hours using Google's public recursive resolver, and in 116~hours using \zdnss own iterative resolver. \zdns achieves 2.6--85~times more successful queries per second and up to 35\% less packet drop compared to existing tools.

\subsection{Performance and Scalability}
\label{sub:sec:perf}

\zdnss performance is dependent on a variety of configuration parameters.
We evaluate \zdnss execution time and success rate (i.e., a \texttt{NOERROR} or \texttt{NXDOMAIN} response) performing \texttt{A} and \texttt{PTR} lookups while modulating available resources. 
We evaluate performance using Cloudflare and Google's public recursive resolvers as well as ZDNS's own iterative resolver. 

We perform all experiments with 24~virtual cores, 16GB of system memory, one process per CPU core (the Go default), and an available 45K~ephemeral ports.
While we do not vary the server specifications for benchmarking, we vary the amount of threads and cache-size that \zdns is exposed to, which can be used as an approximation for measuring the computational resources \zdns requires. 
We find that a single virtual core uses 100\% of resources at approximately 2K~\zdns threads and RAM usage never exceeds 10GB across all cache sizes in our experiments.
Domain names are drawn from a corpus of 234M~fully qualified domain names found on unexpired, browser-trusted certificates in Censys~\cite{censys-2015}; we provide a breakdown of the domains in Appendix~\ref{app:data_breakdown}.  
We do not overlap names or IPv4 addresses between consecutive trials to minimize the impact of  resolver caching. 

\begin{figure*}[h!]
	\centering
	\includegraphics[width=\textwidth]{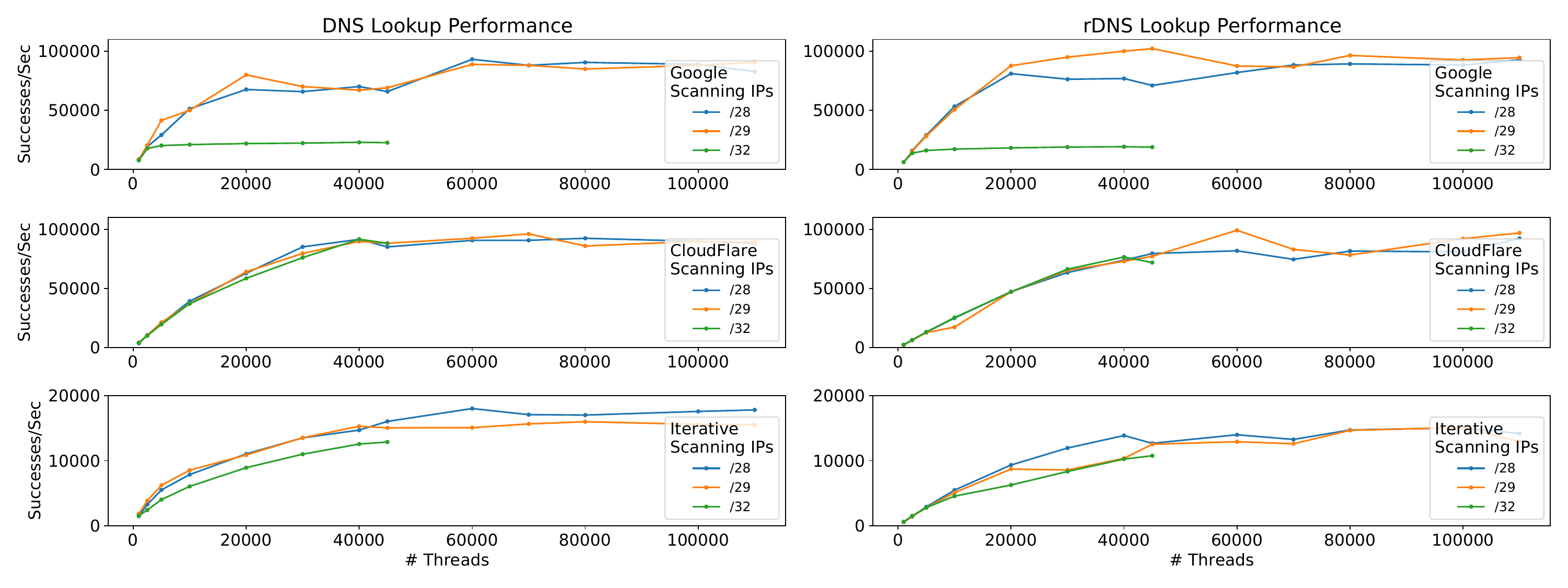}
	\caption{\textbf{\zdns Scalability}---%
	\textnormal{ When using at least 45K threads, \zdnss successfully performs 91.6K A lookups per second when using the Cloudflare resolver, and 102K PTR lookups per second when using Google's resolver. When \zdns uses one IP address (i.e., a /32) to scan from, it encounters a socket limit---thereby not being able spawn more threads---and a Google rate limit that decreases the success rate by a factor of six.   } 
 }
	\label{fig:threads}
\end{figure*}

\vspace{3pt} \noindent \textbf{Results.}\quad
\zdnss performance is directly dependent on choosing the optimal number of threads (i.e., light-weight Go routines), the resolver, and number of client IPs.
We modulate between 1K--100K threads and scanning from a /32, /29, or /28 sub-network of IPs while resolving 10M~domains/IPs in Figure~\ref{fig:threads}. Across resolvers, peak performance plateaus at roughly 50K threads, with \zdns successfully completing 91.6K~A lookups per second when using the Cloudflare resolver, and 102K PTR lookups per second when using Google's resolver. 
The number of available scanning IPs limits the number of threads \zdns can use, as each thread requires its own socket and unique source port to send and receive traffic.  
Furthermore, we experience a Google per-client-IP rate-limiting~\cite{googleRateLimit} when using a single IP client IP address, decreasing the successes rate by a factor of six compared to using Cloudflare's resolver, which does not rate limit clients~\cite{cfrate}.

To respect resolver rate limiting, and to expose the internal characteristics of DNS operations, \zdns performs its own recursive resolution at a peak performance of 18K~lookups per second. 
This performance dip is due to (1) \zdnss inability to leech off of a public resolver's full cache, and (2) the increased number of iterative queries that \zdns must perform to obtain the final response. 
The number of queries that \zdnss iterative resolver sends in order to receive the final record is nearly equivalent to the number of queries and successes/second when using the Google resolver; at 50K threads, \zdns sends 67K queries per second when resolving PTR records for IPv4 addresses compared to Google's 71K successes per second. 

\begin{figure}[h!]
	\centering
	\includegraphics[width=\columnwidth]{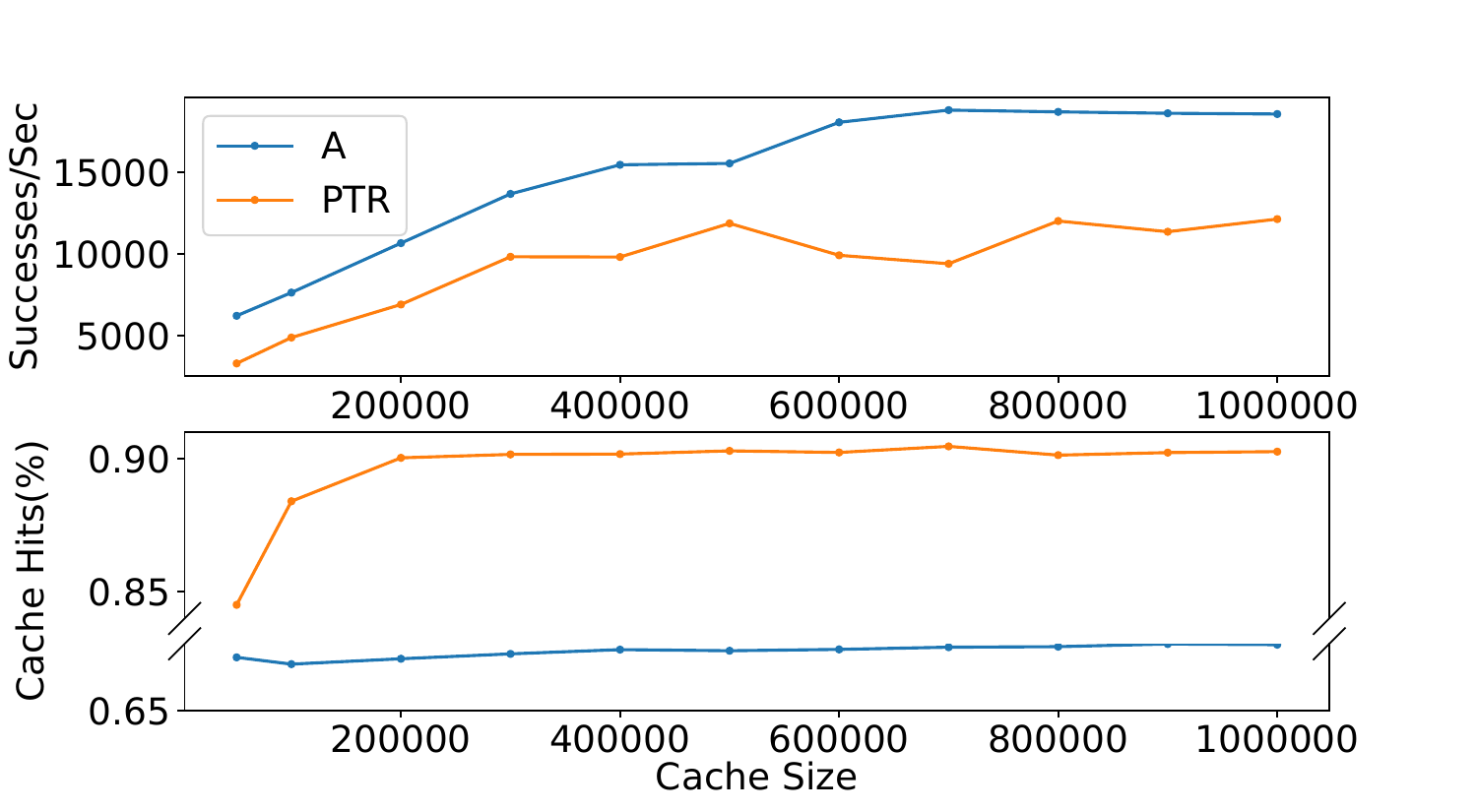}
	\caption{\textbf{\zdns Internal Resolver Cache Performance}---%
	\textnormal{Increasing the cache size results in over 3~times the amount of successes/second.}
 }
	\label{fig:cache}
\end{figure}

\zdnss iterative resolver relies on its own selective caching of responses in order to avoid repeating queries and achieve greater performance. We evaluate \zdnss A and PTR lookup performance while modulating cache size between 50K--1M entries while using 50K~threads. 
Figure~\ref{fig:cache} illustrates that while increasing \zdnss cache size creates a marginal increase (less than 5\%) in cache hit rate, it increases the number of successes per second greater than three-fold. Performance plateaus at a cache size of 600K entries. 


\begin{table}[t]
\centering
\small
\begin{tabular}{llllll}
\toprule
Lookup & Resolver & \# Domains/IPs & \% Successes & Time\\ 
\midrule
A & Google & 50M &96.4\% & 10.6m \\
A & Cloudflare  & 50M & 97.0\% & 10.3m\\
A & Iterative & 50M  & 96.7\% & 46.3m \\
PTR & Google & 100\% IPv4  &93.0\% & 12.1h \\
PTR & Cloudflare &100\% IPv4  &93.5\% & 12.9h \\
PTR & Iterative&100\% IPv4  & 88.5\% &  116.7h  \\
\bottomrule
\end{tabular}
\vspace{3pt}
\caption{
\zdns Performance---%
\textnormal{\zdns resolves 100\% of the IPv4 address space in 12.1~hours using a public recursive resolver and in 116.7~hours using its own iterative resolver.}}
\label{table:perf}
\end{table}
\zdns performance scales when performing large amounts of consecutive lookups (Table~\ref{table:perf}). For example, \zdns resolves the entire IPv4 address space in 12.1~hours when using Google's public resolver and in 116.7~hours when using its own iterative resolver, while maintaining a success rate of 93\% and 88.5\%, respectively.
\zdnss success rate therefore drops less than 2\% when scaling from millions to billions of lookups.

\subsection{Alternative Approaches}

While there are few systems specifically architected for large-scale DNS measurement research, \zdns is not the first system to implement an exposed lookup chain, recursive/caching resolving, or stub resolving. 
We compare \zdns performance against three popular tools---Dig~\cite{dig}, Unbound~\cite{unbound}, MassDNS~\cite{massdns}---when resolving 10M~random IPs/domains with the same server configuration from Section~\ref{sub:sec:perf}.
We configure \zdns to use 60K~threads, a cache-size of 600K~entries, and up to 5~retries per query. 

\vspace{3pt}
\noindent
\textbf{Exposed Lookup Chain.}\quad
Dig~\cite{dig} is a command-line tool used for troubleshooting DNS and provides the ability to ``trace''/expose the full DNS lookup chain by iteratively querying a domain starting from its root nameserver. Dig was never designed to be a high performance scanning engine and we find that its batch lookup performs an average 0.5 A/PTR traces per second. Forking individual dig processes is more efficient and achieves a peak performance of 120~successful A lookups per second when using Cloudflare's resolver to perform a trace query. Beyond its shy performance, Dig's output is not programmatically interpretable, requiring an additional tool for parsing. We compare \zdns and Dig's outputs in Appendix~\ref{app:sub:trace}. 

\vspace{3pt}
\noindent
\textbf{Recursive/Caching Resolving.}\quad
Unbound~\cite{unbound} is a recursive resolver that is commonly used to provide recursive DNS services to clients. To fairly evaluate Unbound's performance using the same resources as \zdnss iterative resolver, 
we install a performance-optimized~\cite{unbound_perf} version of Unbound on \zdnss server\footnote{In practice, Unbound resolvers are often not co-located with the querying program.} and configure \zdns to use the locally installed resolver.
Unbound is less CPU efficient than \zdnss iterative resolver, creating resource contention and capping \zdns to a maximum of 10K and 5K~threads when querying PTR and A records, respectively. 
While ZDNS and Unbound achieve the same number of successes, \zdnss iterative resolver successfully resolves 2.6--3.6~times more names per second (Table~\ref{table:mass_alpha}). 

\vspace{3pt}
\noindent
\textbf{Stub Resolver.}\quad
MassDNS~\cite{massdns} is a high-performance DNS stub resolver that was developed concurrently to ZDNS\@. Unfortunately, during our evaluation, we find that its default behavior overwhelms DNS resolvers, which causes 35\% of responses to either drop or instigate a SERVFAIL (Table~\ref{table:mass_alpha}). To overcome the high failure rate, MassDNS performs up to an additional 50~retries, which further overloads resolvers.
We caution users to approach the tool carefully given the potential to overload servers.



\begin{table}[h]
\centering
\small
\begin{tabular}{lllll}
\toprule
Tool & Lookup & Resolver  & Success/Sec  & \% Total \\ 
& & & &Success\\
\midrule
MassDNS & A & Google  & 197K & 65\% \\
& PTR & Google & 179K & 61\%  \\
& A & Cloudflare & 224K & 67\% \\
& PTR & Cloudflare & 183K& 63\% \\
\midrule
\zdns & A & Unbound & 4.9K & 96\% \\
& PTR & Unbound & 4.5K & 91\%  \\
& A & Iterative &  18K &  97\% \\
& PTR & Iterative &  11.8K &  90\% \\
& A & Google  & 93.1K &  96\% \\
& PTR & Google & 88.8K & 93\%  \\
& A & Cloudflare & 92.5K &  97\% \\
& PTR & Cloudflare & 99.1K &  94\%\\

\bottomrule
\end{tabular}
\vspace{3pt}
\caption{
Alternatives vs \zdns: \textnormal{\zdns's iterative resolver performs 2.6--3.6 more successful queries per second than Unbound, and experiences roughly 30\% less packet drop than MassDNS. }}

\label{table:mass_alpha}
\end{table}


\newcommand{\tableTCP}{
\begin{table}
  \begin{tabularx}{\linewidth}{Xrrr}
      \toprule
      \textbf{Trial} & \textbf{Names} & \textbf{UDP} & \textbf{TCP} \\
      \midrule
      \multirow{4}{*}{10,000 threads} &  84,291,869\,(22\%) &  Failed  &  Success \\
                                      & 118,734,638\,(31\%) &  Failed  &  Failed  \\
                                      &  44,188,444\,(12\%) &  Success &  Failed  \\
                                      & 134,601,895\,(35\%) &  Success &  Success \\
      \midrule
      \multirow{4}{*}{1,000 threads} &   70,092,546\,(18\%) &  Failed  &  Success \\
                                     &  115,125,072\,(30\%) &  Failed  &  Failed  \\
                                     &   63,908,519\,(17\%) &  Success &  Failed  \\
                                     &  132,690,709\,(35\%) &  Success &  Success \\
      \bottomrule
  \end{tabularx}
  \caption{\textbf{TCP vs UDP resolution}\,---\,%
  We resolved all 382,816,846 names from trusted certificates using both TCP
  and UDP, varying the number of threads used across trials. Across both
  trials, 311,109,472~(81.2\%) distinct names successfully resolved in at
  least one trial for at least one transport.
  }
  \label{table:tcp}
\end{table}
}

\section{Case Study: Nameserver (In)consistency}

In this section, we demonstrate the extensibility and utility of \zdnss iterative resolver by analyzing the availability and consistency of redundant nameserver deployment. 
To increase DNS reliability, RFC~1034~\cite{rfc1034} and~2182~\cite{rfc2182} require that a zone must have topologically distributed redundant nameservers. 
Prior work has found, however, that nameservers are not always redundant and vary within their ability to support TCP fallback~\cite{mao2022assessing}, CAA records returned~\cite{ccr2018_caa_first_look}, and parent-child delegations~\cite{sommese2020parents}. 

To compare nameserver responses within the lookup chain, we use \zdnss iterative resolver and add functionality to query and record the responses of all name servers when resolving a domain (i.e., an ``\textit{--all-nameservers}'' flag). 
The functionality is implemented in 30~lines of code.
Using the new functionality, \zdns resolves all 234M fully qualified domain names from our evaluation set (Appendix~\ref{app:data_breakdown}), allowing up to 10~retries for each query to minimize transient errors and approximate the availability of servers.
\zdns completes the scan in 18.5~hours. 

\vspace{3pt} \noindent \textbf{Availability.}\quad
We compare the availability of nameservers for each domain by comparing the number of queries retried across nameservers. 
Nameservers are remarkably consistent in availability; only 0.55\% of resolvable domains have at least one nameserver that needs at least two retries. Interestingly, 0.01\% of domains have at least one nameserver that requires 10~retries in order to elicit a response, 31\% of which belong to the namebrightdns.com domain.
Domains belonging to the Vietnam ccTLD and Nigerian ccTLD are also often unavailable, contributing to 11\% and 7\% of the inconsistent domains, respectively. 
We find no relationship between domain categories (e.g., medical, entertainment) and domain availability when categorizing domains using Cloudflare's DNS categories~\cite{cloud-domains}. 
Upon follow-up probing, we find that the availability inconsistencies are likely not due to server overload, but rather a temporary probabilistic blocking, similar to SSH~\cite{wan2020origin}, in which single, yet consecutive, queries cause a temporary response timeout.  
 \looseness=-1

\vspace{3pt} \noindent \textbf{Response Consistency.}\quad
Over 99.99\% of domains return consistent sets of A records across nameservers, due to the high-centralization of domains being hosted by large response-consistent providers such as Cloudflare (12\%), and GoDaddy (12\%). 
Our results paint a much more consistent picture of DNS compared to prior work~\cite{sommese2020parents}, likely due to the different source of domain names (i.e., CT logs vs. zone files) and the continued centralization of the Internet.


\section{Case Study: CAA Records}
\label{sec:caa}

In this section, we demonstrate \zdnss versatility with alternative record types by analyzing the Certification Authority
Authorization (CAA) ecosystem. CAA records allow a domain owner to specify the
Certificate Authorities (CAs) authorized to issue certificates; the expectation is that a CA will validate the CAA
record before certificate issuance~\cite{rfc6844,rfc8659}.  
Notably, Scheitle~et al.,~\cite{ccr2018_caa_first_look}---the first and primary work to analyze the CAA ecosystem---explicitly advocated for the community to develop an open-source tool capable of querying CAA records. 
Furthermore, while existing data sets~\cite{openintel} query
CAA records on a continuous basis, they exclude the vast majority of ccTLDs due to lack of zone files, which we show contribute to nearly half of all CAA record holders.

\zdns implements CAA records by changing less than five lines of code in the template module shown in Appendix~\ref{app:sub:ex_mod}, and adding 15 lines of additional code tailored to CAA-records across two other framework-specific files.  
We use \zdns to query the CAA records
of all $93M$ base domains from our
evaluation set (Appendix~\ref{app:data_breakdown}), in which 55\% of domains are legacy generic TLDs, 39\% are country code TLDs, and 6\% are new generic TLDs. 
\zdns is able to follow CNAMEs for CAA validation, per RFC~8659~\cite{rfc8659}.\footnote{If \dns{cname.example.com} is a CNAME to \dns{example.net},
then on a request to issue certificate for \dns{cname.example.com} \zdns follows the
CNAME chain to request the CAA records for
\dns{example.net}~\cite{rfc8659,lets_encrypt_caa}.}

\vspace{3pt} \noindent \textbf{CAA Deployment.}\quad
Of the $64M$ domains with a NOERROR response, $1.08M$
domains ($1.69\%$) respond to CAA queries, with 8,000 domains
requiring the CNAME chain to be followed to obtain the CAA record.
ccTLDs are 20\% more likely to hold a CAA record than a gTLD and contribute to 48\% of all CAA records.
The \dns{.pl} ccTLD---a ccTLD absent from most open data sets---alone accounts for 25\% of all CAA-enabled ccTLD domains. 
The top~10 ccTLDs together account for 70\% of all CAA-enabled ccTLD domains.

\vspace{3pt} \noindent \textbf{CAA Configuration.}\quad
The vast majority of the CAA enabled domains configure the tags correctly:
99\% of domains use either or both the \dns{issue} (96.8\%)  and \dns{issuewild} (55.27\%) tags.
While the use of \dns{iodef} tags is generally limited (6.87\% ), 647~domains---many appearing to be affiliated with Visa Inc---use only the \dns{iodef} tag.
Further, we find that 459 domains (0.04\%) are configured with invalid tags. 
We trace the root cause of the majority of these invalid tags to incorrect input validation by a large registrar.
We report the issue to the registrar, prompting them to fix the issue. 

\vspace{3pt} \noindent \textbf{CAA Issuers.}\quad
In 2017, Scheitle~et al.,~\cite{ccr2018_caa_first_look} found that Let's Encrypt~\cite{aas2019letsencrypt} was present in roughly 60\% of CAA records
We find that, five years later---and when considering ccTLDs---Let's Encrypt is present nearly all (92.4\% of \dns{issue} and 93.48\% of \dns{issuewild\xspace}) CAA records. 
Further, Comodo~\cite{comodoca} and Digicert~\cite{digicert} are now present in over 50\% of domains. 


\section{Challenges and Future Work}

While \zdns has enabled efficiently querying large numbers of names, the DNS ecosystem is continually evolving and there are several avenues for future work. This includes extending \zdns support for encrypted DNS lookups, including DNS over HTTPS~(DoH)~\cite{rfc8484} and DNS over TLS~(DoT)~\cite{rfc7858}. Unfortunately, encrypted DNS protocols require \zdns to maintain a TCP connection, eschewing the UDP socket re-use optimization that contributes to \zdns performance (Section~\ref{sec:arch}). 
Furthermore, \zdns will need to integrate a TLS library (e.g., ZCrypto~\cite{zcrypto}) which will cause additional latency when performing cryptographic computations and maintaining stateful TLS connections.  
To maintain \zdnss fast performance, we will explore adding optimizations to \zdnss core architecture such as the integrating the reuse of TLS connections across multiple resolutions. 
Other community-requested features include exposing the capabilities  of \zdns as a library, integrating additional tests, and adding more metadata to returned results.


\section{Conclusion}
\zdns is a modular, extensible, fast, open-source DNS toolkit optimized for quickly and safely performing billions of recursive DNS queries within hours. 
As the domain name system continues to grow in search space, add more record types, and implement more complex functionality, \zdns is built to effortlessly scale and to be easily extended. We hope that \zdns helps 
researchers to better understand, build, and secure the DNS ecosystem. \zdns is released under the Apache 2.0 licence at \url{https://github.com/zmap/zdns}.
\section*{Acknowledgements}
We thank the dozens of open source contributors to the \zdns project, as well as Stefan Savage, Geoffrey Voelker, Gerry Wan, Tatyana Izhikevich, Katherine Izhikevich, Vishal Mohanty, Deepak Kumar, Brian Dickson, Duane Wessels, members of the Stanford University security and networking groups, our shepherd, Fabi\'{a}n E. Bustamante, and the anonymous reviewers for providing insightful discussion and comments. This work was supported in part by the National Science Foundation under award CNS-1823192, Google., Inc., the NSF Graduate Fellowship DGE-1656518, the Office of Naval Research under award N00014-18-1-2662, and a Stanford Graduate Fellowship.
{
\balance
\bibliographystyle{ACM-Reference-Format}
\bibliography{reference}
\appendix
\onecolumn

\section{Evaluation Dataset}
\label{app:data_breakdown}
\begin{wraptable}{r}{6.5cm}
\vspace{-10pt}
\begin{tabular}{lrrr}
    \toprule
    & \multicolumn{1}{l}{fqdn} & \multicolumn{1}{l}{domain} & \multicolumn{1}{l}{tld} \\
    \midrule
    legacy gTLDs & 129644044 & 45865899 & 5 \\
    ngTLDs & 14228236 & 6094090 & 1211 \\
    ccTLDs & 90659109 & 41574286& 486 \\
    \midrule
    All Domains & 234531389 & 93534275 & 1702 \\
    \bottomrule
\end{tabular}
\vspace{5pt}
\caption{Certificate Transparency Domains Dataset---\textnormal{We present the breakdown of domain types used for our evaluations in Sections~\ref{sec:eval} and ~\ref{sec:caa}}}
\label{tab:evalutaion_domain_breakdown}
\end{wraptable}

To evaluate \zdns, we extract all unique fully qualified domain names found on unexpired, browser-trusted certificates in Censys~\cite{censys-2015}.
The 234M fully-qualified~domains (fqdn) found in the Certificate Transparency logs present a good mix of different types of domains: 55\% are legacy generic TLDs, 39\% are country code TLDs,  and  6\% are new generic TLDs (Table~\ref{tab:evalutaion_domain_breakdown}). The 234M fqdns map to 93M base domains. When running a \zdns scan with default parameters, roughly 70\% of the domain names successfully resolve.

\section{Example Module}
\label{app:sub:ex_mod}
Implementing DoLookup functions and modules is simple.
We provide an example DoLookup function in Figure~\ref{fig:doLookup}, and an example module that performs Sender Policy Framework (SPF) lookups in Figure~\ref{fig:module}.

\begin{figure*}[h!]
	\begin{small}
\begin{Verbatim}[obeytabs]
func (s *Lookup) DoLookup(namestring, nameServer string) (interface{}, zdns.Trace, zdns.Status, error) {
	innerRes, trace, status, err := s.DoMiekgLookup(
	                                    miekg.Question{Name: name, Type: s.DNSType, Class: s.DNSClass}
	                                   , nameServer)
	resString, resStatus, err := s.CheckTxtRecords(innerRes, status, err)
	res := Result{Spf: resString}
	return res, trace, resStatus, err
}
\end{Verbatim}
\end{small}
	\caption{\textbf{Example DoLookup}---%
	\textnormal{Implementing DoLookup function is simple, such as this Sender Policy Framework lookup function. } 
 }
 	\label{fig:doLookup}
\end{figure*}


\begin{small}
\begin{Verbatim}[obeytabs]

// SPF Module ================================================================
package spf

import (
    "github.com/zmap/dns"
    "github.com/zmap/zdns/pkg/miekg"
    "github.com/zmap/zdns/pkg/zdns"
    "regexp"
)

const spfPrefixRegexp = "(?i)^v=spf1"

// result to be returned by scan of host
type Result struct {
	Spf string `json:"spf,omitempty" groups:"short,normal,long,trace"`
}

// Per Connection Lookup ======================================================
type Lookup struct {
	Factory *RoutineLookupFactory
	miekg.Lookup
}

func (s *Lookup) DoLookup(name string, nameServer string) (interface{}, zdns.Trace, zdns.Status, error) {
	innerRes, trace, status, err := s.DoMiekgLookup(
	                                    miekg.Question{Name: name, Type: s.DNSType, Class: s.DNSClass}
	                                   , nameServer)
	resString, resStatus, err := s.CheckTxtRecords(innerRes, status, err)
	res := Result{Spf: resString}
	return res, trace, resStatus, err
}
\end{Verbatim}
\end{small}

\begin{figure*}[h!]
	\begin{small}
\begin{Verbatim}[obeytabs]
// Per GoRoutine Factory ======================================================
type RoutineLookupFactory struct {
	miekg.RoutineLookupFactory
	Factory *GlobalLookupFactory
}

func (rlf *RoutineLookupFactory) MakeLookup() (zdns.Lookup, error) {
	lookup := Lookup{Factory: rlf}
	nameServer := rlf.Factory.RandomNameServer()
	lookup.Initialize(nameServer, dns.TypeTXT, dns.ClassINET, &rlf.RoutineLookupFactory)
	return &lookup, nil
}

func (rlf *RoutineLookupFactory) InitPrefixRegexp() {
	rlf.PrefixRegexp = regexp.MustCompile(spfPrefixRegexp)
}

// Global Factory =============================================================
type GlobalLookupFactory struct {
	miekg.GlobalLookupFactory
}

func (s *GlobalLookupFactory) MakeRoutineFactory(threadID int) (zdns.RoutineLookupFactory, error) {
	rlf := new(RoutineLookupFactory)
	rlf.RoutineLookupFactory.Factory = &s.GlobalLookupFactory
	rlf.Factory = s
	rlf.InitPrefixRegexp()
	rlf.ThreadID = threadID
	rlf.Initialize(s.GlobalConf)
	return rlf, nil
}

// Global Registration ========================================================
func init() {
	s := new(GlobalLookupFactory)
	zdns.RegisterLookup("SPF", s)
}

\end{Verbatim}
	\end{small}
	\caption{\textbf{Example Module}---%
	\textnormal{Implementing a module is simple, such as this Sender Policy Framework module. } 
 }
 	\label{fig:module}
\end{figure*}
\clearpage

\section{Exposed Lookup Chain: Dig vs \zdns}
\label{app:sub:trace}

Released prior to \zdns, dig~\cite{dig} is a command-line tool with the ability to ``trace'' (i.e., expose) the DNS lookup chain. 
We configure both dig and \zdns to expose the lookup chain when querying the A record of google.com, and compare their outputs in 
Figure~\ref{fig:dig_trace} and Figure~\ref{fig:zdns_trace}.
\zdns's output is more programmatically interpretable compared to dig.

\begin{figure*}[h!]
\begin{footnotesize}
\begin{Verbatim}[obeytabs]
; <<>> DiG 9.11.3-1ubuntu1.16-Ubuntu <<>> google.com @1.1.1.1 +trace
;; global options: +cmd
.			517042	IN	NS	a.root-servers.net.
.			517042	IN	NS	b.root-servers.net.
.			517042	IN	NS	c.root-servers.net.
.			517042	IN	NS	d.root-servers.net.
.			517042	IN	NS	e.root-servers.net.
.			517042	IN	NS	f.root-servers.net.
.			517042	IN	NS	g.root-servers.net.
.			517042	IN	NS	h.root-servers.net.
.			517042	IN	NS	i.root-servers.net.
.			517042	IN	NS	j.root-servers.net.
.			517042	IN	NS	k.root-servers.net.
.			517042	IN	NS	l.root-servers.net.
.			517042	IN	NS	m.root-servers.net.
.			517042	IN	RRSIG	NS 8 0 518400 20220531170000 20220518160000 47671 . QaVW5itNWx... 

;; Received 1097 bytes from 1.1.1.1#53(1.1.1.1) in 2 ms

com.			172800	IN	NS	i.gtld-servers.net.
com.			172800	IN	NS	f.gtld-servers.net.
com.			172800	IN	NS	b.gtld-servers.net.
com.			172800	IN	NS	l.gtld-servers.net.
com.			172800	IN	NS	d.gtld-servers.net.
com.			172800	IN	NS	c.gtld-servers.net.
com.			172800	IN	NS	j.gtld-servers.net.
com.			172800	IN	NS	e.gtld-servers.net.
com.			172800	IN	NS	h.gtld-servers.net.
com.			172800	IN	NS	g.gtld-servers.net.
com.			172800	IN	NS	a.gtld-servers.net.
com.			172800	IN	NS	k.gtld-servers.net.
com.			172800	IN	NS	m.gtld-servers.net.
com.			86400	IN	DS	30909 8 2 E2D3C916F6DEEAC73294E8268FB5885044A833FC5459588F4A9184CF C41A5766
com.			86400	IN	RRSIG	DS 8 1 86400 20220531170000 20220518160000 47671 . HBLBAX50zT...

;; Received 1198 bytes from 199.9.14.201#53(b.root-servers.net) in 8 ms

google.com.		172800	IN	NS	ns2.google.com.
google.com.		172800	IN	NS	ns1.google.com.
google.com.		172800	IN	NS	ns3.google.com.
google.com.		172800	IN	NS	ns4.google.com.
CK0POJMG874LJREF7EFN8430QVIT8BSM.com. 86400 IN NSEC3 1 1 0 - CK0Q1GIN43N1ARRC9OSM6QPQR81H5M9A  NS SOA RRSIG DNSKEY NSEC3PARAM
CK0POJMG874LJREF7EFN8430QVIT8BSM.com. 86400 IN RRSIG NSEC3 8 2 86400 20220523042356 20220516031356 37269 com. tnQdPZZqo2... 

S84BKCIBC38P58340AKVNFN5KR9O59QC.com. 86400 IN NSEC3 1 1 0 - S84BUO64GQCVN69RJFUO6LVC7FSLUNJ5  NS DS RRSIG
S84BKCIBC38P58340AKVNFN5KR9O59QC.com. 86400 IN RRSIG NSEC3 8 2 86400 20220524051956 20220517040956 37269 com. BQwb7CiufG...

;; Received 836 bytes from 192.35.51.30#53(f.gtld-servers.net) in 60 ms

google.com.		300	IN	A	142.250.188.14
;; Received 55 bytes from 216.239.38.10#53(ns4.google.com) in 48 ms
\end{Verbatim}
\end{footnotesize}
	\caption{\textbf{dig +trace Output}---%
	\textnormal{When exposing the lookup chain, dig's output is not programmatically interpretable, requiring an additional tool for parsing. } }
 	\label{fig:dig_trace}
\end{figure*}

\pagebreak

\begin{figure*}[h!]
\begin{footnotesize}
\begin{Verbatim}[obeytabs]
{ "class": "IN",
  "data": {
    "answers": [{ "answer": "216.58.195.78", "class": "IN", "name": "google.com", "ttl": 300, "type": "A" }],
    "flags": { "authenticated": false, "authoritative": true, "checking_disabled": false, "error_code": 0, "opcode": 0, 
               "recursion_available": false, "recursion_desired": false, "response": true, "truncated": false },
    "protocol": "udp", 
    "resolver": "216.239.34.10:53"
  },
  "name": "google.com",
  "status": "NOERROR",
  "timestamp": "2022-05-18T19:19:58Z",
  "trace":[{"cached": false, "class": 1, "depth": 1, "layer": ".", "name": "google.com", "name_server": "199.7.83.42:53",
            "results": {
                "additionals":[ 
                    {"answer":"192.55.83.30","class":"IN","name":"m.gtld-servers.net","ttl":172800,"type":"A"},
                    {"answer":"192.41.162.30","class":"IN","name":"l.gtld-servers.net","ttl":172800,"type":"A"},
                    {"answer":"192.52.178.30","class":"IN","name":"k.gtld-servers.net","ttl":172800,"type":"A"},
                    ...
                    {"answer":"192.26.92.30","class":"IN","name":"c.gtld-servers.net","ttl":172800,"type":"A"},
                    {"answer":"192.33.14.30","class":"IN","name":"b.gtld-servers.net","ttl":172800,"type":"A"},
                    {"answer":"192.5.6.30","class":"IN","name":"a.gtld-servers.net","ttl":172800,"type":"A"},
                    {"answer":"2001:501:b1f9::30","class":"IN","name":"m.gtld-servers.net","ttl":172800,"type":"AAAA"} ],
                "authorities":[
                    {"answer":"f.gtld-servers.net.","class":"IN","name":"com","ttl":172800,"type":"NS"},
                    {"answer":"d.gtld-servers.net.","class":"IN","name":"com","ttl":172800,"type":"NS"},
                    {"answer":"i.gtld-servers.net.","class":"IN","name":"com","ttl":172800,"type":"NS"},
                    ... 
                    {"answer":"g.gtld-servers.net.","class":"IN","name":"com","ttl":172800,"type":"NS"},
                    {"answer":"c.gtld-servers.net.","class":"IN","name":"com","ttl":172800,"type":"NS"},
                    {"answer":"k.gtld-servers.net.","class":"IN","name":"com","ttl":172800,"type":"NS"} ]
                "flags": { "authenticated": false, "authoritative": false, "checking_disabled": false, "error_code": 0, "opcode": 0, 
                          "recursion_available": false, "recursion_desired": false, "response": true, "truncated": false},
                "protocol": "udp",
                "resolver": "199.7.83.42:53"},
            "try": 1,
            "type": 1 },
           {"cached": false, "class": 1, "depth": 2, "layer": "com", "name": "google.com", "name_server": "192.5.6.30:53",
            "results": {
                "additionals":[
                    {"answer":"2001:4860:4802:34::a","class":"IN","name":"ns2.google.com","ttl":172800,"type":"AAAA"},
                    {"answer":"216.239.34.10","class":"IN","name":"ns2.google.com","ttl":172800,"type":"A"},
                    {"answer":"2001:4860:4802:32::a","class":"IN","name":"ns1.google.com","ttl":172800,"type":"AAAA"},
                    {"answer":"216.239.32.10","class":"IN","name":"ns1.google.com","ttl":172800,"type":"A"},
                    {"answer":"2001:4860:4802:36::a","class":"IN","name":"ns3.google.com","ttl":172800,"type":"AAAA"},
                    {"answer":"216.239.36.10","class":"IN","name":"ns3.google.com","ttl":172800,"type":"A"},
                    {"answer":"2001:4860:4802:38::a","class":"IN","name":"ns4.google.com","ttl":172800,"type":"AAAA"},
                    {"answer":"216.239.38.10","class":"IN","name":"ns4.google.com","ttl":172800,"type":"A"}],
                "authorities":[
                    {"answer":"ns2.google.com.","class":"IN","name":"google.com","ttl":172800,"type":"NS"},
                    {"answer":"ns1.google.com.","class":"IN","name":"google.com","ttl":172800,"type":"NS"},
                    {"answer":"ns3.google.com.","class":"IN","name":"google.com","ttl":172800,"type":"NS"},
                    {"answer":"ns4.google.com.","class":"IN","name":"google.com","ttl":172800,"type":"NS"}],
                "flags": { "authenticated": false, "authoritative": false, "checking_disabled": false, "error_code": 0, "opcode": 0, 
                           "recursion_available": false, "recursion_desired": false,"response": true, "truncated": false },
                "protocol": "udp",
                "resolver": "192.5.6.30:53" },
            "try": 1,
            "type": 1 },
           {"cached": false, "class": 1, "depth": 3, "layer": "google.com", "name":   "google.com",  "name_server": "216.239.34.10:53",
            "results": {
            "answers": [{ "answer": "216.58.195.78", "class": "IN", "name": "google.com", "ttl": 300, "type": "A"}],
            "flags": { "authenticated": false, "authoritative": true, "checking_disabled": false, "error_code": 0, "opcode": 0,
                   "recursion_available": false, "recursion_desired": false, "response": true, "truncated": false },
            "protocol": "udp",
            "resolver": "216.239.34.10:53" },
            "try": 1,
            "type": 1
            }]
}

\end{Verbatim}
\end{footnotesize}
	\caption{\textbf{\zdns \texttt{+trace} Output}---%
	\textnormal{When exposing the lookup chain, \zdns' output is in JSON, which is programmatically interpretable. } 
	}
 	\label{fig:zdns_trace}
\end{figure*}
}

\end{document}